\documentclass[12pt]{article}
\usepackage{epsfig}
\usepackage{epsfig}
\usepackage{psfrag}
\usepackage{cite}
\usepackage{graphicx}
\usepackage{rotate}
\usepackage{latexsym}
\usepackage{amssymb}
\newcommand\pubnumber{
}

\def\csumb{Department of Physics, New York University, \\ 4 Washington Place, New York, NY 10003, USA}

\def\Title#1{\begin{center} {\Large\bf #1 } \end{center}}
\def\Author#1{\begin{center}{ \sc #1} \end{center}}
\def\Address#1{\begin{center}{ \it #1} \end{center}}

\newcommand\pubblock{\rightline{\begin{tabular}{l} \pubnumber\\
\end{tabular}}}
\newenvironment{Abstract}{\begin{quotation}  }{\end{quotation}}

\makeatletter
\def\section{\@startsection{section}{0}{\z@}{5.5ex plus .5ex minus
 1.5ex}{2.3ex plus .2ex}{\large\bf}}
\def\subsection{\@startsection{subsection}{1}{\z@}{3.5ex plus .5ex minus
 1.5ex}{1.3ex plus .2ex}{\normalsize\bf}}
\def\subsubsection{\@startsection{subsubsection}{2}{\z@}{-3.5ex plus
-1ex minus  -.2ex}{2.3ex plus .2ex}{\normalsize\sl}}

\renewcommand{\@makecaption}[2]{%
   \vskip 10pt
   \setbox\@tempboxa\hbox{\small #1: #2}
   \ifdim \wd\@tempboxa >\hsize     % IF longer than one line:
       \small #1: #2\par          %   THEN set as ordinary paragraph.
     \else                        %   ELSE  center.
       \hbox to\hsize{\hfil\box\@tempboxa\hfil}
   \fi}
 \def\citenum#1{{\def\@cite##1##2{##1}\cite{#1}}}
\def\citea#1{\@cite{#1}{}}
\newcount\@tempcntc
\def\@citex[#1]#2{\if@filesw\immediate\write\@auxout{\string\citation{#2}}\fi
  \@tempcnta\z@\@tempcntb\m@ne\def\@citea{}\@cite{\@for\@citeb:=#2\do
    {\@ifundefined
       {b@\@citeb}{\@citeo\@tempcntb\m@ne\@citea\def\@citea{,}{\bf }\@warning
       {Citation `\@citeb' on page \thepage \space undefined}}%
    {\setbox\z@\hbox{\global\@tempcntc0\csname b@\@citeb\endcsname\relax}%
     \ifnum\@tempcntc=\z@ \@citeo\@tempcntb\m@ne
       \@citea\def\@citea{,}\hbox{\csname b@\@citeb\endcsname}%
     \else
      \advance\@tempcntb\@ne
      \ifnum\@tempcntb=\@tempcntc
      \else\advance\@tempcntb\m@ne\@citeo
      \@tempcnta\@tempcntc\@tempcntb\@tempcntc\fi\fi}}\@citeo}{#1}}
\def\@citeo{\ifnum\@tempcnta>\@tempcntb\else\@citea\def\@citea{,}%
  \ifnum\@tempcnta=\@tempcntb\the\@tempcnta\else
  {\advance\@tempcnta\@ne\ifnum\@tempcnta=\@tempcntb \else\def\@citea{--}\fi
    \advance\@tempcnta\m@ne\the\@tempcnta\@citea\the\@tempcntb}\fi\fi}
\makeatother
\newcommand{\bmath}{\begin{displaymath}}
\newcommand{\emath}{\end{displaymath}}
\newcommand{\be}{\begin{equation}}
\newcommand{\ee}{\end{equation}}
\newcommand{\bea}{\begin{eqnarray}}
\newcommand{\eea}{\end{eqnarray}}

\newcommand{\bne}{\bar{\nu}_e}
\newcommand{\eg}{\varepsilon_\gamma}
\newcommand{\lm}{\lambda_{\mbox{{\tiny min}}}}
\newcommand{\nn}{\nonumber}
\begin{document}
\begin{titlepage}
\pubblock
\vfill
\def\thefootnote{\fnsymbol{footnote}}
\boldmath \Title{Radiative Correction to the $\bar{\nu}_e$ ($\nu_e$)
Spectrum in $\beta$-Decay} \unboldmath
%
%\vfill
\Author{
A.~Sirlin$^1$\footnote{\tt{alberto.sirlin@nyu.edu}}}
\Address{$^1$\csumb}
%\vfill
\vspace*{2cm}
\begin{Abstract} 
We derive an analytic expression for the ${\mathcal O}(\alpha)$ radiative correction to the $\bar{\nu}_e (\nu_e)$ spectrum in allowed $\beta$-decay.
The $m \to 0$ limit is convergent and leads to a very simple result ($m$ is the electron mass). This is in sharp contrast  to the correction to the $e^- (e^+)$
spectrum, that diverges as $m \to 0$, an important difference that we explain on theoretical grounds. After discussing some of their general properties, we use the corrections to the $\bar{\nu}_e$ and $e^-$ spectra to derive the corresponding correction to the conversion from the $e^-$ spectrum to the $\bar{\nu}_e$ spectrum, a relation that is of considerable interest for reactor studies of neutrino oscillations. 
\end{Abstract}

%PACS Classification: 11.10.-z; 11.15.Bt; 12.38.Bx; 02.90.+p
\vfill
\end{titlepage}
\def\thefootnote{\arabic{footnote}}
\setcounter{footnote}{0}
%--

%--
\section{Introduction}
The ${\mathcal O}(\alpha)$ radiative correction (R.~C.) to the $e^- (e^+)$ spectrum has played an important role in precision studies of $\beta$-decay.
It is described by a R.~C. $(\alpha/2\pi)g (E,E_0)$, where $E$ and $E_0$ are the energy and end-point energy of the $e^- (e^+)$ \cite{r1,r2}.
Here we have in mind the R.~C. not contained in Fermi's Coulomb function $F(Z,E)$, which is separately included in the theory of $\beta$-decay. The effect of $(\alpha/2\pi)g (E,E_0)$ on the spectrum shape is significant, particularly for large values of $E_0$ \cite{r1,r3}. It also gives important contributions to the theoretical calculations of the Ft-values of superallowed $(0^+ \to 0^+)$ Fermi transitions \cite{r4} and the neutron lifetime \cite{r5}. We recall that the superallowed Fermi transitions play a crucial role in the precise verification of the unitarity of the Cabibbo-Kobayashi-Maskawa (CKM) matrix.
In fact, the current analysis, which includes large short-distance effects \cite{r4}, the R.~C. $(\alpha/2\pi)g (E,E_0)$, and a number of more recent refinements \cite{r5,r6}, leads to the conclusion \cite{r7}:
\be \label{eq1}
|V_{ud}|^2 + |V_{us}|^2 +|V_{ub}|^2 = 0.99995(61)\, ,
\ee 
an impressive test of the three-generation Standard Theory at the level of its quantum corrections. It is also interesting to note that the overall electroweak corrections to Eq.~(\ref{eq1}) are of ${\mathcal O}(4 \%)$, i.~e. 66 times larger than the $0.061 \, \%$ error.

The correction $(\alpha/2\pi)g$, in conjunction with the short-distance effects and refinements discussed in Refs.~\cite{r4,r5,r6}, has also been recently applied to the calculation of the muon capture rate in the singlet state of H \cite{r8}.

The aim of the present paper is to derive the corresponding R.~C. to the $\bne (\nu_e)$ spectrum in allowed $\beta$-decay which, as explained below, is of considerable current interest.

An important difference with the R.~C. to the $e^- (e^+)$ spectrum involves the inner bremsstrahlung (I.~B.) contributions. In fact, the integration over the $e^- (e^+)$ momenta introduces a complicated dependence on the radiated photon energy that hinders the subsequent integration over the photon momenta. In Section~\ref{secII} we develop and implement a convenient strategy to deal with this problem. The combination of the I.~B. and virtual corrections leads then to the R.~C. $(\alpha/2 \pi) h(\hat{E}, E_0)$ to the $\bne (\nu_e)$ spectrum,
which we present in analytic form. Here $\hat{E} \equiv E_0 -K$, where $K$ is the 
$\bne (\nu_e)$ energy. We find that in the $m \to 0 $ limit ($m$ is the electron mass), $h(\hat{E}, E_0)$ converges and has a very simple form. This is in sharp contrast with $g(E, E_0)$, that diverges as $m \to 0$, an important difference that we explain on theoretical grounds.

In Section~\ref{secIII} we apply the R.~C. $(\alpha/2\pi)g (E,E_0)$ and 
$(\alpha/2 \pi) h(\hat{E}, E_0)$ to the conversion from the $e^-$ spectrum to the 
$\bne$ spectrum, a relation that is of considerable current interest for reactor studies of neutrino oscillations~\cite{r9mep,r9,r9bis}. This leads to the analytic expression for the ${\mathcal O}(\alpha)$
R.~C. to that important relation, which we illustrate with numerical examples.

\boldmath
\section{Radiative Correction to the $\bar{\nu}_e$ ($\nu_e$) Spectrum \label{secII}}
\unboldmath

As in Ref.~\cite{r1}, we first consider the basic process of neutron $\beta$-decay, $n \to p + e^- +\bne$, and neglect terms of ${\mathcal O} (\alpha E/M)$, where $M$ is the nucleon mass.

The  ${\mathcal O} (\alpha)$ virtual R.~C., as well as the zeroth order transition probability, are obtained from those in the $e^-$ spectrum by simply replacing $E \to \hat{E}, p \to \hat{p}, \beta \to \hat{\beta}$, where $E, p$, and $m$ are the $e^-$ energy, momentum and mass, $\beta = p/E$, $K$ is the $\bne$
energy and $\hat{E}, \hat{p}$, and $\hat{\beta}$ are defined by
\be \label{eq2}
\hat{E} \equiv E_0 -K \, ; \quad
\hat{p} \equiv \left(\hat{E}^2 - m^2 \right)^{1/2} \, ; \quad \hat{\beta} \equiv
\hat{p}/\hat{E}\, .
\ee 
The reason for this result is that, in deriving the $\bne$ spectrum, one integrates over the $e^-$ phase space factor $d^3p/2 E$ rather than  the $\bne$'s $d^3K/2 K$. In these two contributions the conservation of energy is expressed by $\delta(E_0 - K -E)$ and the phase space integrations lead to the relation explained above. 

We now turn our attention to the inner bremsstrahlung (I.~B.) contributions, which present more subtle and theoretically interesting differences with the $e^-$ spectrum.

Calling $\eg$ the radiated photon energy, the I.~B. amplitude involves two terms: the first one behaves as $1/\eg$ as $\eg \to 0$; the second one is of ${\mathcal O}(\eg^0)$. The square of the absolute value of the first term, combined with the $d^3 k/2 \eg$ photon phase space factor ($k$ is the photon momentum)
leads to the infrared (I.~R.) divergent contribution. The interference of the two terms and the square of the absolute value of the second one are I.~R. convergent and are evaluated separately.

As in \cite{r1}, we regularize the I.~R. divergence by introducing an infinitesimal photon mass $\lambda_{\mbox{{\tiny min}}}$, so that $\eg = (k^2 +\lm^2)^{1/2}$. After integrating over the proton and electron momenta, as well as the $\bne$ direction, the I.~R. contribution is proportional to 
$\int_0^{k_m} (d^3k/2 \eg^3) f(\eg,v)$, where 
\be \label{eq3}
f(\eg,v) = p E\left\{(1/\beta v) \ln{\left[(1+\beta v)/(1-\beta v)\right]} - 
(1-\beta^2)/(1-\beta^2 v^2) -1 \right\}\, ,
\ee
$v = k/\eg$, $k_m = \hat{E} -m$ is the maximum photon energy for given $K$, and it is understood that 
\be \label{eq4}
E = \hat{E} -\eg\, ,
\ee
which follows from energy conservation.

The $\eg$ variable in $f(\eg,v)$ denotes the $\eg$ dependence contained in $E$ via Eq.~(\ref{eq4}) and, therefore, also in $p$, $\beta$, and Eq.~(\ref{eq3}). 
The complicated dependence of $f(\eg,v)$ on $\eg$ and $v$ hinders the subsequent integration over the photon momenta. To deal with this problem, a convenient strategy is to split $f(\eg,v)$ according to
\be \label{eq5}
f(\eg,v) = f(0,v) + \left[f(\eg,v) - f(0,v)\right] \, .
\ee
The I.~R. divergence arises from $f(0,v)$, a function that only involves the photon momenta through its dependence on $v$. Changing the integration variable from $k$ to $v$, the integral  $\int_0^{k_m} (d^3 k /2 \eg^3) f(0,v)$ can be performed analytically and leads to a function of $\hat{\beta}$ and $\ln(k_m/\lm)$. Its combination with the virtual ${\mathcal O}(\alpha)$ R.~C. is well defined since, as expected, the I.~R. divergences cancel exactly.

The integration of the second term in Eq.~(\ref{eq5}) is I.~R. convergent, so we set $\lm =0$ ($v=1$), and replace $f(\eg,v) - f(0,v) \to f(k,1) - f(0,1)$. In this latter expression, the $\eg =k$ dependence is contained in $E$
and, therefore, also in $p$, $\beta$, and $f(k,1)$  (cf. Eqs.~(\ref{eq3}, \ref{eq4})).

Changing the variable of integration from $k$ to $E$ via Eq.~(\ref{eq4}), the integral $\int d^3k/ 2 k^3 [f(k,1) - f(0,1)]$ can be also performed analytically. In particular, we note that the result of this integration 
includes the integral 
\be \label{eq6}
I  = \int_{m/\hat{E}}^1 dt \ln{(1-t)} / (t^2 - m^2/\hat{E}^2)^{1/2} \, , 
\ee
for which we have derived the compact analytic representation
\be
I = - \mbox{Li}_2 \left(2 \hat{\beta}/(1+\hat{\beta}) \right) + \mbox{tanh}^{-1}
\hat{\beta} \ln{\left(2 \hat{\beta}^2/(1+\hat{\beta}) \right)}\, , 
\ee
where
\be \label{eq8}
\mbox{Li}_2 (z) = -\int_0^z (dt/t) \ln(1-t) 
\ee
is the dilogarithm.

As mentioned before, the $d^3 k$ integration of the two remaining contributions, namely the interference of the two terms in the I.~B. amplitude and the absolute value square of the second, are I.R. convergent. Changing again the variable of integration from $k$ to $E$ via Eq.~(\ref{eq4}), they can also be performed analytically.

Combining all  the ${\mathcal O}(\alpha)$ contributions discussed above, we obtain the radiatively corrected expression for the $\bne$ spectrum:
\be \label{eq9}
d P_\nu = d P^0_\nu \left[1 +(\alpha/2 \pi) h(\hat{E}, E_0) \right] \, ,
\ee 
where
\be \label{eq10}
d P^0_\nu =  A\, \hat{p}\, \hat{E}\, F(Z,\hat{E})\, K^2 dK
\ee
is the zeroth order spectrum,
\bea
h(\hat{E},E_0) &=& 3 \ln{\left(\frac{m_p}{m}\right)} +\frac{23}{4} -\frac{8}{\hat{\beta}} \mbox{Li}_2 \left(\frac{2 \hat{\beta}}{1+\hat{\beta}} \right)+
8 \left(\frac{\mbox{tanh}^{-1} \hat{\beta}}{\hat{\beta}} -1 \right) \ln\left(
\frac{2 \hat{E} \hat{\beta}}{m} \right) \nn \\
&& + \,4\, \frac{\mbox{tanh}^{-1} \hat{\beta}}{\hat{\beta}} \left[ \frac{7 +3 \hat{\beta}^2}{8} - 2 \,\mbox{tanh}^{-1} \hat{\beta} \right]\, , \label{eq11}
\eea
$m_p$ is the proton mass, and $\hat{E}, \hat{p}, \hat{\beta}, \mbox{Li}_2(z)$
are defined in Eqs.~(\ref{eq2},\ref{eq8}). In Eq.~(\ref{eq10}), $F(Z,\hat{E})$ is the Fermi Coulomb function evaluated at $E = \hat{E}$ and $A$ is a constant independent of $K$. In neutron $\beta$-decay, for example, $A = (g_V^2+3 g_A^2)/2 \pi^3$, where $g_V = G_F V_{ud}$ and $g_A$ are the vector and axial vector coupling constants.
In analogy with the $e^-$ spectrum case~\cite{r1}, Eqs.~(\ref{eq9}-\ref{eq11}) can be applied to allowed nuclear
$\beta$ decay using an independent particle model of the nucleus. In that case $g_V^2+3 g_A^2$ is replaced by
$g_V^2 |M_F|^2+ g_A^2 |M_{GT}|^2$, where $M_F$ and $M_{GT}$ are the Fermi and Gamow-Teller matrix elements. In analogy with 
$g(E,E_0)$,  Eq.~(\ref{eq11}) is valid for both $\bar{\nu}_e$ and ${\nu}_e$ transitions.

Aside from the corrections $(\alpha/2 \pi) g(E, E_0)$ and $(\alpha/2 \pi) h(\hat{E}, E_0)$, in the Standard
Theory there are large additional electroweak corrections which give very important contributions to the
Ft-values of the superallowed Fermi transitions and  the neutron lifetime. However, they are constants independent 
of $K$ and $E$, so they don't affect the shape of the $\bar{\nu}_e$ and $e^-$ spectra. For this reason,
they are not included in the present discussion.

Eq.~(\ref{eq11}) simplifies considerably in the $m \to 0$ ($\hat{\beta} \to 1$) limit. Noting that
$\mbox{tanh}^{-1} \hat{\beta} = \ln{\left[ \hat{E} (1+\hat{\beta})/m \right]}$ and $\mbox{Li}_2 (1) = \pi^2 / 6$, we readily find:
\be \label{eq12}
\lim_{m \to 0} h(\hat{E},E_0) = 3 \ln{\left(\frac{m_p}{2 E_0}\right)} +\frac{23}{4} -\frac{4 \pi^2}{3}- 3 \ln{(1-y)}\, ,
\ee
where $y=K/E_0$. Thus, the  $m \to 0$ limit of $h(\hat{E}, E_0)$ is convergent and has a very simple form. This is in
sharp contrast with the behavior of $g(E,E_0)$, the function that describes the R.~C. to the $e^- (e^+)$ spectrum. In fact,
in the relativistic $m^2 << E^2$ limit

\bea
g(E,E_0) & \to &  3 \ln{\left(\frac{m_p}{2 E_0}\right)} -\frac{3}{4} -\frac{2 \pi^2}{3} \nn \\
&& + 4  \left(\ln{x} - 1 \right) \left[ \frac{1-x}{3x} -\frac{3}{2}+\ln{\left(\frac{1-x}{x}\right)} \right] +
 \frac{(1-x)^2}{6x^2} \ln{x} \nn \\ 
&& + \ln{\left(\frac{2 E_0}{m}\right)} \left[ \frac{4(1-x)}{3x}-3+\frac{(1-x)^2}{6x^2} +4 \ln{\left(\frac{1-x}{x}\right)} \right]    \, ,           \label{eq13}
\eea
where  $x=E/E_0$. Eq.~(\ref{eq13}) can be gleaned from the expression between curly brackets in Eq.~(4.4) of Ref.~\cite{r3},
after subtracting $ 6 \ln{(\lambda/m_p)} +9/4$ (cf. footnote 23 in Ref.~\cite{r1}).
Thus, the correction to the electron spectrum diverges logarithmically in the $m \to 0$ limit! On the other hand,
if one multiplies Eq.~(\ref{eq13}) by $x^2(1-x)^2$, which describes the energy dependence of the zeroth order probability
in the  $m \to 0$ limit, and integrates over $0 \le x \le 1$, the cofactor of $\ln{(2 E_0/m)}$ vanishes, so that the correction
to the inverse lifetime is indeed free from the electron ``mass singularity''~\cite{r3,r10,r11}. 

The fact that  $h(\hat{E}, E_0)$ is convergent in the  $m \to 0$ limit while $g(E,E_0)$ is not, can be explained 
by the following theoretical argument. For given $K$, in the $m \to 0$ limit all collinear $e-\gamma$ configurations
become energy degenerate with energy $E_0-K$ and generally give rise to mass singularities (cf. Eq.~(\ref{eq3})). An
elementary, but powerful theorem in Quantum Mechanics leads to the conclusion that these singularities are cancelled
in the power series expansions of transition probabilities if the latter are summed over an appropriate ensemble
 of such degenerate states~\cite{r11}. In deriving the corrections to the $\bar{\nu}_e (\nu_e)$ spectrum, we perform the
$d^3 p$ and  $d^3 k$ integrations, so indeed we sum over the set of collinear  $e-\gamma$ configurations that 
become energy degenerate in the $m \to 0$ limit. Thus, according to this theorem,  $h(\hat{E}, E_0)$ should be
free of $\ln m$ singularities, as we find.
In contrast, this is not the case in the derivation of the electron spectrum, since the $d^3 p$ integration is not carried out.
As a consequence, $g(E,E_0)$ (Eq.~(\ref{eq13})) explicitly exhibits a $\ln{m}$ singularity. Analogous examples
of mass singularities in the differential spectrum, and their cancellation in the lifetime, integrated asymmetry 
and some partial decay rates in muon decay are extensively discussed in Ref.~\cite{r3}.

A useful consistency check of Eqs.~(\ref{eq12},\ref{eq13}) is to multiply Eq.~(\ref{eq12}) by $y^2(1-y)^2$, 
Eq.~(\ref{eq13}) by $x^2(1-x)^2$, and integrate over $x$ and $y$ from 0 to 1. The two results
should be equal since both lead to the correction to the inverse lifetime, a fact that we have verified
to high accuracy.  Eq.~(\ref{eq12}) can also be derived directly by neglecting the terms in the integrands
 that vanish in the $m \to 0$ limit. In that approximation the phase space integrations are very simple,
and we find that the $\ln{m}$ singularities from the various contributions indeed cancel, leading again to Eq.~(\ref{eq12}).

As an illustrative example, we consider the case $E_0 = 8$ MeV. Using  Eq.~(\ref{eq11}), we find that the R.~C. 
$(\alpha/2 \pi) h(\hat{E}, E_0)$ to the  $\bar{\nu}_e (\nu_e)$ spectrum increases the transition probability by 1.17 \% at
$y=0.8$ ($K=6.4$ MeV) and by  0.64 \% at $y=0.2$ ($K=1.6$ MeV). This behavior is reflected in the simpler expression in 
Eq.~(\ref{eq12}), which shows a positive, monotonically increasing function of $y$. Using the exact expression for
$g(E,E_0)$ (Eq.~(20b) of Ref.~\cite{r1}), we find that the R.~C.  $(\alpha/2 \pi) g(E, E_0)$ to the  $e^- (e^+)$ spectrum
decreases the transition probability by  1.13 \% at
$x=0.8$ ($E=6.4$ MeV) and increases it by 2.90 \% at $x=0.2$ ($E=1.6$ MeV).
This behavior can be qualitatively understood by noting that the cofactor of $\ln{(2 E_0/m)}$ in Eq.~(\ref{eq13})
is negative for $x \gtrsim 0.44$ and positive for $x \lesssim 0.44$. As a consequence, the R.~C. decreases (increases) the transition probability for
high (low) energy  $e^-$ or  $e^+$, an effect that becomes sharper as  $E_0$ increases. A similar behavior has been
found in muon decay~\cite{r3}.

In many cases, Eqs.~(\ref{eq12},\ref{eq13}) provide very good and useful approximations to the exact expressions
for  $h(\hat{E}, E_0)$ and $g(E, E_0)$. Their domains of validity are $(E_0-K)^2 >> m^2$ and $E^2 >> m^2$, respectively.
As shown in Ref.~\cite{r1}, the  ${\mathcal O}(\alpha)$  R.~C. to the shape of the  $e^- (e^+)$ spectrum 
is not affected by strong interactions if terms of  ${\mathcal O}(\alpha E/M)$ are neglected. The same conclusion applies 
to the ${\mathcal O}(\alpha)$  R.~C. to the $\bar{\nu}_e (\nu_e)$ spectrum.

\boldmath
\section{Radiative Correction to the Conversion from the $e^-$ Spectrum to the $\bar{\nu}_e$ Spectrum \label{secIII}}
\unboldmath

Including the ${\mathcal O}(\alpha)$  R.~C., the theoretical expressions for the $e^-$ and  $\bar{\nu}_e$ spectra
in allowed $\beta$-decay can be cast in the form
\bea
\frac{d P_e}{d E} &=& f_e(E, E_0)\, , \label{eq14}\\
\frac{d P_\nu}{d K} &=&  f_\nu(K, E_0)\, , \label{eq15}
\eea
where
\bea
f_e(E, E_0) &=&  A\, p\, E\, (E_0-E)^2\, F(Z, E)\, \left[1+ \frac{\alpha}{2 \pi} g(E, E_0) \right]\, , \label{eq16}\\
f_\nu(K, E_0) &=&  A\, \hat{p}\, \hat{E}\, K^2\, F(Z, \hat{E})\, \left[1+ \frac{\alpha}{2 \pi} h(\hat{E}, E_0) \right]\, , \label{eq17}
\eea
and $\hat{E}$ and  $\hat{p}$ are defined in  Eq.~(\ref{eq2}). Comparing  Eqs.~(\ref{eq16},\ref{eq17})
and recalling that $\hat{E} = E_0-K$, we see that 
\be \label{eq18}
f_\nu(K, E_0) = f_e(\hat{E}, E_0) \left[1+ \frac{\alpha}{2 \pi} h(\hat{E}, E_0) \right] 
/ \left[1+ \frac{\alpha}{2 \pi} g(\hat{E}, E_0) \right] \, ,
\ee
or, neglecting terms of order  ${\mathcal O}(\alpha^2)$: 
\be \label{eq19}
f_\nu(K, E_0) = f_e(\hat{E}, E_0) \left[1+ \frac{\alpha}{2 \pi} \left( h(\hat{E}, E_0) - g(\hat{E}, E_0) \right) \right]\, .
\ee
Using Eq.~(\ref{eq11}) and the exact expression for $g(E,E_0)$ (Eq.~(20b) of Ref.~\cite{r1}), we find
\bea
& & h(\hat{E}, E_0) - g(\hat{E}, E_0) = \frac{13}{2} -\frac{4}{\hat{\beta}} \mbox{Li}_2 
\left(\frac{2 \hat{\beta}}{1+\hat{\beta}} \right) \nn \\
& & + 4 \left(\frac{\mbox{tanh}^{-1} \hat{\beta}}{\hat{\beta}} -1 \right) \left[\frac{11}{6}+ \ln\left(
\frac{2 E_0}{m} \right) + 2 \ln{\hat{\beta}}+ 2 \ln{(1-y)} - \ln{y} -\frac{1}{3(1-y)}\right] \nn \\
& & + \frac{\mbox{tanh}^{-1} \hat{\beta}}{\hat{\beta}} \left[\frac{3-\hat{\beta}^2}{2}- 4 \,\mbox{tanh}^{-1} \hat{\beta}
- \frac{y^2}{6 (1-y)^2}\right]\, , \label{eq20}
\eea
where $\hat{\beta}$ is defined in Eq.~(\ref{eq2}) and we recall that  $y \equiv K/E_0$.

 Eqs.~(\ref{eq19},\ref{eq20}) describe the conversion from the $e^-$ spectrum in a specific decay into the
corresponding  $\bar{\nu}_e$ spectrum when 
 ${\mathcal O}(\alpha)$  R.~C. are included.

As an illustrative example, we consider again the case  $E_0 = 8$ MeV. Using  Eqs.~(\ref{eq19},\ref{eq20}) we find that, 
for a given $e$ spectrum $f_e(E, E_0)$, the  R.~C. $(\alpha/2 \pi) \left[ h(\hat{E}, E_0) - g(\hat{E}, E_0) \right]$
decreases $f_\nu(K, E_0)$ by 1.74 \% at $y=0.8$ ($K=6.4$ MeV) and increases it by 1.77 \% at $y=0.2$ ($K=1.6$ MeV).
Thus, the correction factor $[1 + (\alpha/2 \pi) (h(\hat{E},E_0) - g(\hat{E},
E_0)]$ in
Eq.~(\ref{eq19}) is $3.57\, \%$ larger at $K = 1.6 \, \mbox{MeV}$
 than at $K = 6.4\,  \mbox{MeV}$.
This behavior reflects the dominant contribution of $- g(\hat{E}, E_0)$. As a consequence, the  R.~C. 
decreases (increases)  $f_\nu(K, E_0)$ for high (low) $K$, an effect that becomes sharper as  $E_0$ increases.

\section{Conclusions}

In summary, we have derived an analytic expression for the ${\mathcal O}(\alpha)$ R.~C. to the  $\bar{\nu}_e (\nu_e)$ spectrum
in allowed $\beta$-decay (Eqs.~(\ref{eq9}-\ref{eq11})). We have found that the $m \to 0$ limit of this correction is
convergent and leads to a very simple expression (Eq.~(\ref{eq12})). This is in sharp contrast to
the ${\mathcal O}(\alpha)$ R.~C. to the $e^- (e^+)$ spectrum, that diverges as $m \to 0$, an important difference
that we have explained on theoretical grounds. We have then used our results to derive the ${\mathcal O}(\alpha)$ R.~C. to the
conversion from the $e^-$ spectrum to the $\bar{\nu}_e$ spectrum (Eqs.~(\ref{eq19},\ref{eq20})), a relation that plays an important r\^ole in
reactor studies of neutrino oscillations~\cite{r9mep,r9,r9bis}.
A recent study concludes that the conversion procedure can be implemented with an error  $\lesssim 1 \%$, provided some conditions are met \cite{r9}.
If this accuracy is achieved, it is likely that the R.C. will play a significant role in precision studies of the $\bar{\nu}_e$ spectrum.

\section*{Acknowledgments} \noindent
The author is indebted to W.~J.~Marciano for calling his attention to the conversion from the $e^-$ spectrum to the $\bar{\nu}_e$ spectrum, and for very useful observations, and to A.~Ferroglia and G.~Ossola for very interesting and helpful discussions. This work was supported in part by NSF Grant No. PHY-0758032.


\begin{thebibliography}{99}
\bibitem{r1}
  A.~Sirlin,
  %``General Properties of the Electromagnetic Corrections to the Beta Decay of a Physical Nucleon,''
  Phys.\ Rev.\  {\bf 164}, 1767-1775 (1967).



\bibitem{r2}
  A.~Sirlin,
  Acta \ Physica \ Austriaca, \ Supplement {\em V}, 
  ``Particles, Currents, Symmetries'', edited by P. Urban, p. 353-390 (Springer-Verlag, New York, 1968).

\bibitem{r3}
  T.~Kinoshita, A.~Sirlin,
  %``Radiative corrections to Fermi interactions,''
  Phys.\ Rev.\  {\bf 113}, 1652-1660 (1959).

\bibitem{r4}
  A.~Sirlin,
  %``Current Algebra Formulation of Radiative Corrections in Gauge Theories and the Universality of the Weak Interactions,''
  Rev.\ Mod.\ Phys.\  {\bf 50}, 573-605 (1978), Erratum-ibid. {\bf 50}, 905 (1978). 
  
\bibitem{r5}
  A.~Czarnecki, W.~J.~Marciano, A.~Sirlin,
  %``Precision measurements and CKM unitarity,''
  Phys.\ Rev.\  {\bf D70 } 093006 (2004).
%  [hep-ph/0406324].


\bibitem{r6}
  W.~J.~Marciano, A.~Sirlin,
  %``Improved calculation of electroweak radiative corrections and the value of V(ud),''
  Phys.\ Rev.\ Lett.\  {\bf 96}, 032002 (2006).
%  [hep-ph/0510099].



\bibitem{r7}
  J.~C.~Hardy, I.~S.~Towner,
  %``Superallowed 0+ ---> 0+ nuclear beta decays: A New survey with precision tests of the conserved vector current hypothesis and the standard model,''
  Phys.\ Rev.\  {\bf C79}, 055502 (2009).
%  [arXiv:0812.1202 [nucl-ex]].




\bibitem{r8}
  A.~Czarnecki, W.~J.~Marciano, A.~Sirlin,
  %``Electroweak radiative corrections to muon capture,''
  Phys.\ Rev.\ Lett.\  {\bf 99}, 032003 (2007).
%[arXiv:0704.3968 [hep-ph]].

\bibitem{r9mep} C. Bemporad, G. Gratta, P. Vogel, Rev. Mod. Phys. {\bf 74}, 297 (2002), and
references cited therein.

\bibitem{r9} P. Vogel, Phys. Rev. {\bf C76}, 025504 (2007), and references cited therein.

\bibitem{r9bis} Th. A. Mueller et al., [arXiv:1101.2663 [hep-ex]], and references cited therein.


\bibitem{r10}
 T.~Kinoshita,
  %``Mass singularities of Feynman amplitudes,''
  J.\ Math.\ Phys.\  {\bf 3}, 650-677 (1962).
  

\bibitem{r11}
  T.~D.~Lee, M.~Nauenberg,
  %``Degenerate Systems and Mass Singularities,''
  Phys.\ Rev.\  {\bf 133}, B1549-B1562 (1964).
  


\end{thebibliography}
\end{document}